\documentclass[aps,prl,twocolumn,showpacs,superscriptaddress,groupedaddress]{revtex4}

\usepackage{graphicx}
\usepackage{color}
\usepackage{dcolumn}
\usepackage{bm}
\usepackage{amssymb}
\usepackage{amsmath}
\usepackage{textcomp}
\usepackage{natbib}

\begin{document}

\title{Electromagnetic Burst Generation during Annihilation of Magnetic Field in Relativistic Laser-Plasma Interaction}

\author{Y.~J. Gu}
\email{yanjun.gu@eli-beams.eu}
\affiliation{Institute of Physics of the ASCR, ELI-Beamlines, Na Slovance 2, 18221 Prague, Czech Republic}
\affiliation{Institute of Plasma Physics of the CAS, Za Slovankou 1782/3, 18200 Prague, Czech Republic}
\author{F. Pegoraro}
\affiliation{Enrico Fermi Department of Physics, University of Pisa, Pisa, Italy}
\affiliation{National Research Council, National Institute of Optics, (CNR/INO ), Pisa, Italy}
\author{P. V. Sasorov}
\affiliation{Institute of Physics of the ASCR, ELI-Beamlines, Na Slovance 2, 18221 Prague, Czech Republic}
\affiliation{Keldysh Institute of Applied Mathematics of the Russian Academy of Sciences, Miusskaya Ploshchad' 4, Moscow, 125047, Russia}
\author{D. Golovin}
\affiliation{Institute of Laser Engineering, Osaka Univesity, 2-6 Yamada-Oka, Suita, Osaka, 565-0871 Japan}
\author{A. Yogo}
\affiliation{Institute of Laser Engineering, Osaka Univesity, 2-6 Yamada-Oka, Suita, Osaka, 565-0871 Japan}
\affiliation{PRESTO, Japan Science and Technology Agency, Kawaguchi, Saitama 332-0012, Japan}
\author{G. Korn}
\affiliation{Institute of Physics of the ASCR, ELI-Beamlines, Na Slovance 2, 18221 Prague, Czech Republic}
\author{S. V. Bulanov}
\affiliation{Institute of Physics of the ASCR, ELI-Beamlines, Na Slovance 2, 18221 Prague, Czech Republic}
\affiliation{Kansai Photon Research Institute, National Institutes for Quantum and Radiological Science and Technology, 8-1-7 Umemidai, Kizugawa-shi, Kyoto 619-0215, Japan}
\affiliation{Prokhorov General Physics Institute of the Russian Academy of Sciences, Vavilov Str. 38, 119991 Moscow, Russia}

\date{\today}

\begin{abstract}
%We present the results of theoretical studies of formation and evolution of the current sheet in a colliosionless plasma
%during magnetic reconnection in relativistic limit. Relativistic magnetic reconnection is driven by parallel laser pulses interacting with underdense plasma target. Annihilation of laser created magnetic field of opposite polarity generates strong non-stationary electric field formed in between the region with opposite polarity magnetic field accelerating charged particles within the current sheet. This laser-plasma target configuration is discussed in regard with the laboratory modeling of charged particle acceleration and gamma flash generation in astrophysics.
We present the results of 3-dimensional kinetic simulations and theoretical studies on the formation and evolution of the current sheet in a collisionless plasma during magnetic field annihilation in the ultra-relativistic limit.
Annihilation of oppositively directed magnetic fields driven by two laser pulses interacting with underdense plasma target is accompanied by an electromagnetic burst generation. The induced strong non-stationary longitudinal electric field accelerates charged particles within the current sheet. Properties of the laser-plasma target configuration
are discussed in the context of the laboratory modeling for charged particle acceleration and gamma flash generation in astrophysics.
\end{abstract}

%\pacs{52.27.Ny, 52.38.Ph, 52.35.Kt, 52.65.Rr}
%\pacs{52.27.Ny, 52.35.Vd, 52.38.Fz, 52.65.Rr}
%\keywords{}

\maketitle

%Introduction
%\section{Introduction}
%\label{intro}
\section{I. Introduction}

%Magnetic reconnection (MR) is one of the fundamental problems in plasma physics. %Motivated by explaining the acceleration mechanism in space plasmas, the idea of MR, which allows the energy transfer from magnetic field into charged particles, is proposed.
Magnetic reconnection (MR) is a fundamental process in astrophysical and laboratory plasmas which provides a mechanism to convert the magnetic field energy to the energy of plasmas and charged particles accompanying with the topology variation of the magnetic field lines \cite{BiskampMR, AstrophysBookBulanov, PriestMR}. Various impulsive phenomena in space plasmas can be attributed to MR such as solar flares \cite{Parker1957, Lin2003, SuNature2013}, coronal mass ejections \cite{Jiong2004, Fermo2014}, pulsar winds \cite{Kirk2001, Kuijpers2015, Cerutti2017}, open and closed planetary magnetospheres \cite{PhysRevLett.16.1207, Brady2009, Faganello2012} and gamma-ray bursts \cite{Giannios2010, Bing2011, McKinney2012, Benoit2012}. In laboratory plasmas, MR is related to the nuclear fusion plasma dynamics \cite{doi:10.1063/1.1706761, White:2204327, Yamada1994, Hastie1997} and the weakly ionized plasmas \cite{BulanovSakai1998}.

The observations of high-energy gamma rays from the Crab Nebula reported by \textit{Agile} and \textit{Fermi-LAT} in 2010 is tightly related to MR \cite{Abdo739, Tavani736, refId0, GammaRayMR2014}. According to Ref. \cite{0004-637X-749-1-26}, the underlying mechanism of gamma-flares in the Crab Nebulas is based on the electron acceleration during magnetic reconnection in the relativistic limit. The features of the time dependence of gamma-flares indicate that the electron acceleration occurs in non-stationary electric field. Modeling of such the phenomena under the conditions of terrestrial laboratories is an intriguing problem.

The dynamics of laser produced plasmas has been shown in Ref. \cite{Askaryan1995, Bulanov2009, BulanovPPR2015, Hoshino2018} to provide a framework where MR can be investigated in the regimes of interest for astrophysical environment.
%Motivated by the development of high power lasers, the laser-plasma interactions have been employed to investigate MR in the regimes of interest for astrophysical environment.
%\cite{Askaryan1995, Bulanov2009}.
One of the pioneering experiments on MR studying in laser-plasmas was proposed by Nilson \textit{et al.} \cite{PhysRevLett.97.255001} with two nanosecond laser beams and a planar solid target. Li \textit{et al.} \cite{PhysRevLett.99.055001} irradiated a thin plastic (CH) foil with two or four 1-ns long OMEGA laser beams. Both experiments observed the plasma jets with keV energy in the reconnection region. Recent experimental works with intense and high-energy laser pulses have shown plasma outflows with keV electrons and plasmoid generation in current sheets formed during reconnection on the time scale of nanoseconds \cite{Zhong2010}. Fan-like plasma outflows with MeV electrons and a plasmoid were obtained in a similar setup with Al foils by Dong \textit{et al.} \cite{PhysRevLett.108.215001}. Lezhnin \textit{et al.} presented the results of the magnetohydrodynamics simulations of driven magnetic reconnection on colliding magnetized laser-produced plasmas \cite{doi:10.1063/1.5044547}.
%By using the petawatt and short pulse laser, the MR transits into the collisionless relativistic regime. Great deal of studying these effects have been done to conduct computer simulations in relativistic regime.
In the presence of a very powerful (petawatt) short laser pulse MR transits into the collisionless relativistic regime. Several numerical studies have been performed in the relativistic regime.
Ping \textit{et al.} \cite{PhysRevE.89.031101, 0004-637X-849-2-137} reported a fast reconnection driven by two ultra-intense laser pulses using 3D kinetic simulations and the corresponding change of the topology structure was observed. MR driven by nonthermal and relativistic electron beams have been discussed recently in Refs. \cite{Hoshino2001, FiuzaPRL2016, PhysRevE.98.043207}. Magnetic reconnections under the extreme condition of QED critical field are proposed in Refs. \cite{2018arXiv180709750S, 2018arXiv180910772H}.

In ultra-relativistic regime, the MR acquires novel features. Due to the relativistic constraint on the electron velocity, the variation of the magnetic field cannot be sustained by the upper limit of the electron current. It is principaly dominated by the displacement current and the corresponding inductive electric field. This so-called dynamic dissipation of the magnetic field was first proposed by S. I. Syrovatskii \cite{Syrovatskii1966}. Recent numerical studies with particle-in-cell (PIC) simulations presented clear signatures of the particle acceleration in the magnetic annihilation regime \cite{PhysRevE.93.013203, :/content/aip/journal/pop/22/10/10.1063/1.4933408, gu_TEM10_2016, 0741-3335-60-4-044020}.

In this paper, we present the results of kinetic simulations on the collisionless relativistic MR regime in the 3D configurations. The purpose is to investigate the electron acceleration via the MR generated electric field. We consider the configuration in which two sub-petawatt short laser pulses interact with the hydrogen plasma target comprising two density steps. The magnetic fields with opposite polarities generated by the laser driven electron current is expected to annihilate in the low density region due to the transverse expansion of the magnetic field. The fast annihilation
creates strong electric field accelerating electrons up to high energy. The corresponding topology variation of the magnetic field lines is observed.

The paper is organized as follows. In Section II we describe the simulation setup. Section III presents the results of the kinetic simulations showing two electron current filaments generated by two laser pulses interacting with underdense plasma target and the associated magnetic field configuration produced by these electron current filaments.
Section IV is devoted to the description of opposite magnetic polarities merging resulting in the magnetic X-line formation and its evolution to thin current sheet. In Section V
we discuss the excitation of the electromagnetic burst and the corresponding charged particle acceleration. The dynamics of the electrons and the typical trajectories are presented. Reconnection of
the magnetic field lines in collisionless plasmas is closely related to the Hall effect resulting in the transverse electric field and current excitation. The Hall effect and the quadruple magnetic field formed in the vicinity of the X-line are considered in Section VI. In Section VII we describe the tearing mode-like instability leading to the current
sheet break-up into magnetic islands. Section VIII contains discussions and conclusions.

\section{II. Simulation Setup}

The kinetic simulations are based on the relativistic electromagnetic code EPOCH \cite{Ridgers2014273, 0741-3335-57-11-113001} in a 3-dimensional configuration.
%\re{PIC simulation is a widely used and reliable method to investigate the dynamics in plasmas. In this method, physical particles are represented by a number of pseudoparticles. The fields generated by the laser pulse and the motion of particles are calculated by a Finite Difference Time Domain (FDTD) method. All the electromagnetic field components are calculated within a grid with fixed spatial resolution. The forces generated by these fields are applied on the pseudoparticles and used to update their velocities and positions. At the end of the loop, the new calculated pseudoparticles� positions and velocities are used to update the fields again.}
The simulation box has the size of $L_x=240~\lambda$ and $L_y=L_z=90~\lambda$. Here $\mathrm{\lambda=1~\mu m}$ is the laser wavelength. The mesh size in the simulations is $\delta x=\delta y=\delta z=\lambda /20$. All the quasiparticles (8 per cell) are initially at rest with a total number of $7\times10^7$. The real mass ratio of electron and proton ($m_p/m_e=1836$) is used in the simulations. Open boundary conditions are applied for both fields and particles. Two linearly polarized Gasussian pulses with the peak intensity of $\mathrm{10^{21}~W/cm^2}$ propagating in parallel along the $x$-axis are focused on the plane $x=15~\lambda$. The optical axes of the two pulses are transversely separated by a distance equal to $30~\lambda$, being located at $y=\pm15\lambda$. The normalized amplitude is $a_0=eE_0/m_e\omega c \approx 27$, where $E_0$ and $\omega$ are the laser electric field strength and frequency, $e$ and $m_e$ are the electron charge and mass, respectively; and $c$ is speed of light in vacuum. The pulse duration is $\tau=30$~fs and the spot size (FWHM) is of $5~\lambda$. The near critical density hydrogen plasma target has a thickness of $220~\lambda$ in the x-direction and remains uniform in the transverse direction in the region of $\sqrt{y^2+z^2}<35\lambda$. The density linearly increases from 0 to $\mathrm{n_1=0.2n_c}$ in the region $10\lambda<x<15\lambda$ and then remains constant for $40~\lambda$. Here $n_c=m_e \omega ^2 /4\pi e^2$ is the plasma critical density, which is approximately $10^{21}$cm$^{-3}$
for $1\mu$m wavelength laser radiation.  From $x=55~\lambda$ to $165~\lambda$, the density decreases to $\mathrm{n_2=2\times10^{-3}n_c}$. This second density plateau region extends over the length of $60~\lambda$ from $x=165~\lambda$ to $225~\lambda$. By employing a density downramp region, the magnetic field is forced to expand in the lateral direction quickly as discussed below in Section IV. The second density plateau suppresses the strength of the longitudinal electric field arrising due to the electric charge separation effect so that the inductive electric field effect can be clearly distinguished.

\section{III. Electron Current Filaments and Strong Magnetic Field Generation}

The laser pulses generate plasma channels in the underdense hydrogen target since the power of the pulses is higher than the relativistic self-focusing threshold, $P>P_cn_c/n_0$, where $P_c=2m_e^2c^5/e^2=17\mathrm{GW}$. Balancing the electron energy gain from the charge separation field and the laser field, one obtains the radius of the plasma channel as: $R_{ch}=\sqrt{a_{ch}\,n_c/n_0}({\lambda}/{\pi})$, where $\mathrm{a_{ch}}$ is the amplitude of the laser pulse vector potential in the channel \cite{PhysRevSTAB.18.061302}. Due to the self-focusing effect of the laser field, $a_{ch}\propto({P\, n}/{P_c\, n_c})^{1/3}$ becomes higher than the initial dimensionless amplitude $a_{0}$. The corresponding channel radius in this case is $R_{ch}\approx5\mu m$, which is well consistent with the simulation results as shown in Fig. 1(a). The electron density distribution (at $\mathrm{t=110~T_0}$, $\mathrm{T_0=\lambda/c}$ is the laser period) on the bottom plane of Fig. 1(a) presents a double channel structure. In each channel, there is an electron beam trapped and accelerated by the wakefield. The strength of the wakefield is given by Sprangle \textit{et al.} \cite{Sprangle1988} as: $E_{max}\approx0.76\sqrt{n_0}{a_0}/{\sqrt{1+a_0^2}}$. The maximum amplitude reaches $\mathrm{50~GV/cm}$ so that the trapped electrons can be accelerated to high energy in a short distance. The energy spectra of the total electrons at $90~T_0$, $100~T_0$ and $110~T_0$ are plotted in the (y,z) plane. The corresponding energy spectra of the high energy electrons, which are trapped and accelerated by the wakefield, are presented in the (x,z) plane. The peak energy increases from $\mathrm{75~MeV}$ to $\mathrm{125~MeV}$ within about $20~\mu m$. Each accelerated beam contains a large charge $\mathrm{3nC}$ of the relativistic electrons ($\mathrm{\cal{E}}_{ke}>1MeV$).

Fig. 1(b) presents the electron density distribution in the 3D space. The cloud represents the walls structure of the plasma channels. The two accelerated high charge electron beams (the bunches inside the channel) generate the parallel currents, which are shown in the bottom plane. The laser intensity distribution is projected on the (y,z) plane. Due to the self-focusing effect, the intensity becomes as large as two times of the initial peak intensity $I_0$. The currents produce magnetic fields according to Ampere-Maxwell law. The z-component of the azimuthal field ($B_z$) is projected in the x-y plane, which displays two dipole structures (the positive and negative polarities generated by one electron beam). The strength of the magnetic field can be estimated from Ampere-Maxwell law,
\begin{equation}
\label{eq:Mxwll}
c\nabla \times \mathbf{B}=4\pi\mathbf{j}+ \partial_t \mathbf{E}.
\end{equation}
 By assuming a quasistatic condition of $\partial_t \mathbf{E}=0$, one obtains that $B\approx 4\pi neR_{ch}$, which reaches about $\mathrm{0.5~GG}$.

\begin{figure*}
\begin{center}
\resizebox{90mm}{!}{\includegraphics{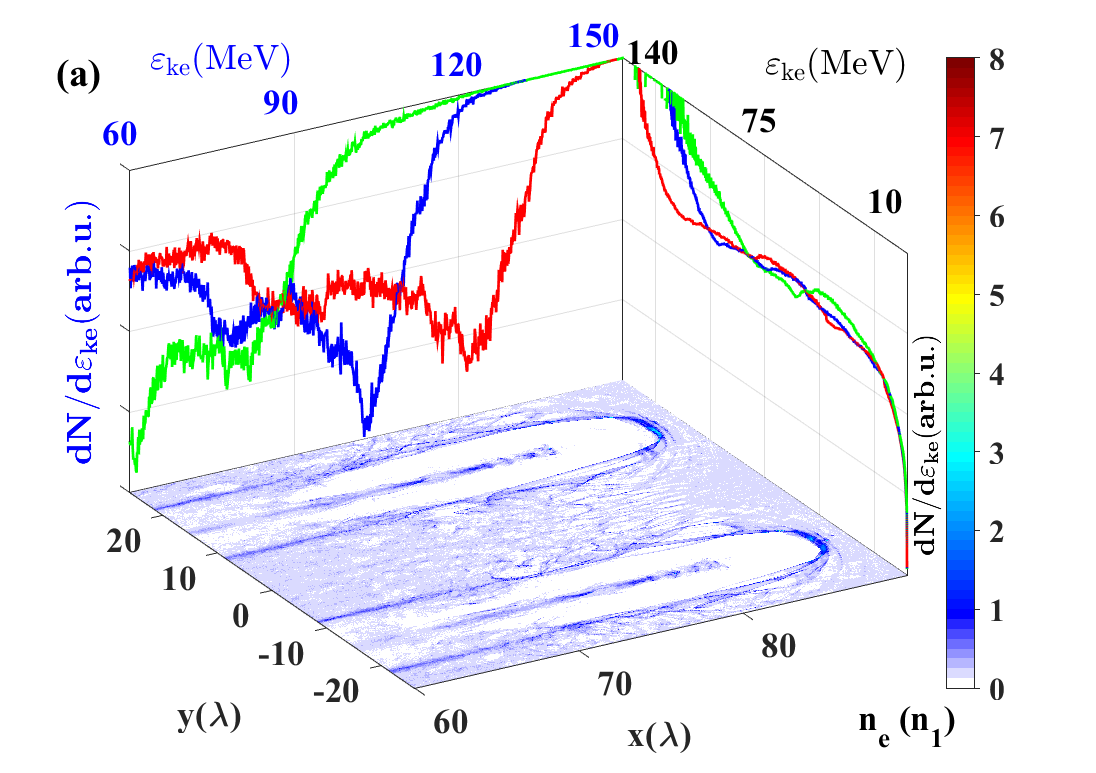}}
\resizebox{80mm}{!}{\includegraphics{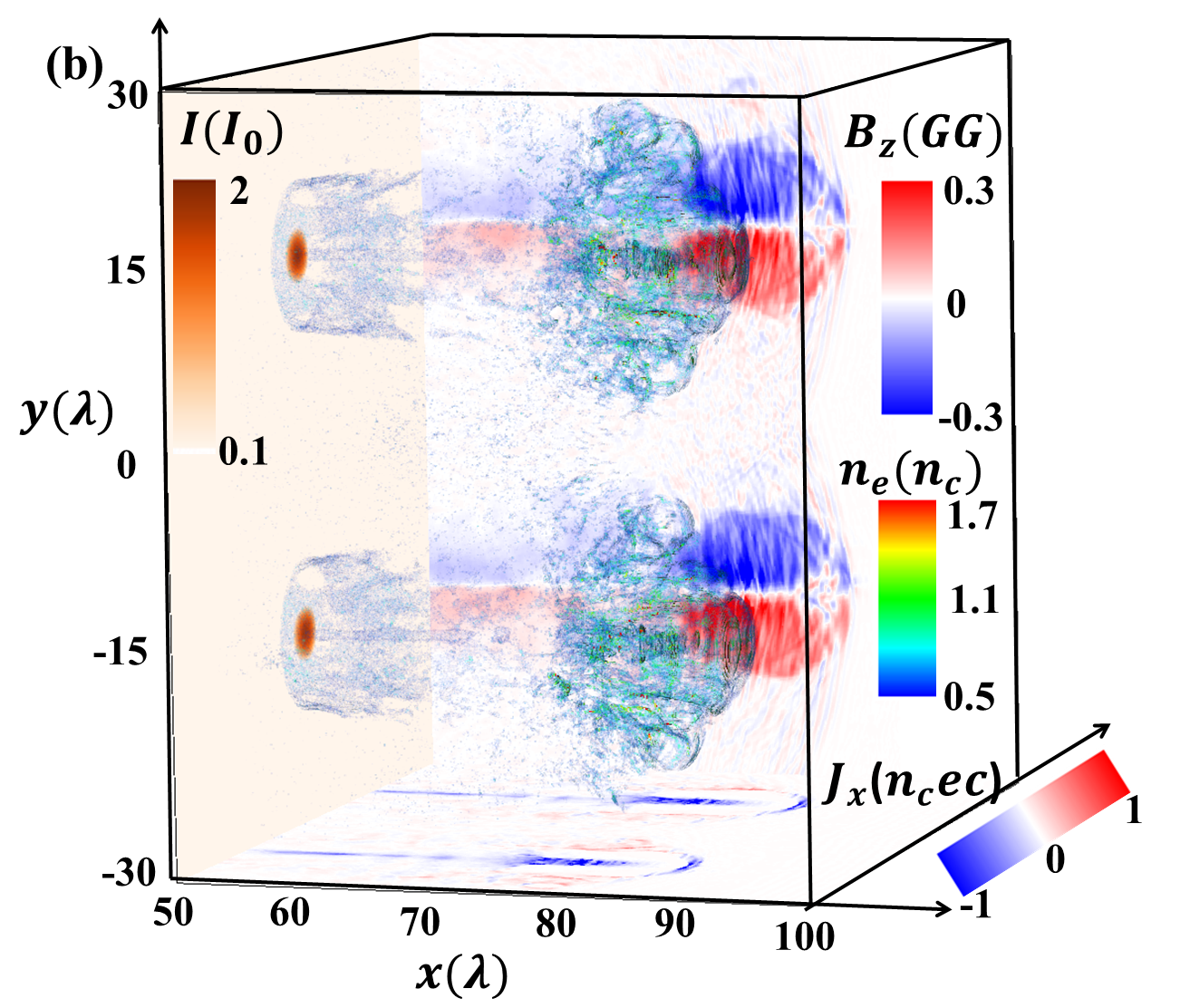}}
%\resizebox{80mm}{!}{\includegraphics{Electron0038AngularEnergy.png}}
%\resizebox{160mm}{!}{\includegraphics{CurretnXBVCX127To132T0038RR10.png}}
\caption{
(color online)
(a) The electron density distribution (for the plane cut at z=0) at $t=110T_0$ is plotted in the bottom plane. In the (y,z) plane, the curves represent the total energy spectrum at $90~T_0$ (green), $100~T_0$ (blue) and $110~T_0$ (red). The energy spectra of the wakfield accelerated electrons are shown in the (x,z) plane with the corresponding time. (b) The transparent cloud represents the electron density distribution in 3D. The current density (along z=0 plane) is projected in the bottom. The distribution of the $B_z$ (along z=0 plane) is projected. The laser intensity distribution (in the center of the pulse) is projected on the (y,z) plane.
}
\label{Fig1}
\end{center}
\end{figure*}

\section{IV. X-line Formation in the Process of Opposite Magnetic Polarities Merging}

The radius of the magnetic dipole structure depends on the radius of the plasma channel $R_{ch}$. In Fig. 2(a), the distribution of the longitudinal components of the current density ($j_x$) is shown ( zooming in the region of the upper half of the simulation box). The magnetic field generated by the forward accelerated electrons ($j_e$) is shielded by the return electron currents ($j_r$) which move along the channel wall. The return currents provide a magnetic field with opposite direction in the region out of the channel. As mentioned in Section III, the radius of the channel is a function of the local plasma density, $R_{ch}=\sqrt{a_{ch}n_c/n_0}({\lambda}/{\pi})$ and $a_{ch}\propto({P n}/{P_c n_c})^{1/3}$. Therefore, the channel expands with the downramp of the density according to $R_{ch}\propto n^{-1/3}$. The expansion velocity depends on the gradient of the density as $\dot{R_{ch}}\propto-v_x\nabla n$, where $v_x$ is the forward propagating velocity of the channel structure, i.e. the group velocity of the laser pulse propagating in the plasma.

With the expansion of the channels, the size of the magnetic dipoles also increase in the transverse direction. As illustrated in Fig. 2(b), the return currents shift from the red ones ($j_r$) to the green ones ($j_r^{\prime}$). The black curves represent the amplitude of the magnetic field generated by the accelerated electron beam ($j_e$), i.e. the amplitude of the z-component of the field ($\mid B_z \mid$). The dashed parts are shielded before the expansion of the channel. Due to the outward shift of the return currents, the dashed parts also amnifest. The contour lines for the magnetic field on the (y,z) plane also shows the expansion from the solid to the dashed region.

The $B_z$ distribution in the plane of $z=0$ at $\mathrm{t=152~T_0, 174~T_0}$ and $\mathrm{196~T_0}$ are shown in Fig. 2(c), (d) and (e). It's clear that the boundaries of the magnetic field extend from $20~\lambda$ to $30~\lambda$, while the strength of the magnetic field decreases. It is due to the outward shift of the return currents reducing their contribution to the magnetic field. The curves describe the profiles of $B_z$, in which the gradient around $y=0$ reflects the variation of the magnetic field (arising from the expansion and collision of opposite magnetic polarities). In Fig. 2(c) at $152~T_0$, the magnetic fields with opposite polarity do not overlap each other in the center and the curve is smooth. However, with the propagation in the density downramp, the opposite polarities start to interacting and the curve become sharp as seen in Fig. 2(e).

\begin{figure*}
\begin{center}
\resizebox{80mm}{!}{\includegraphics{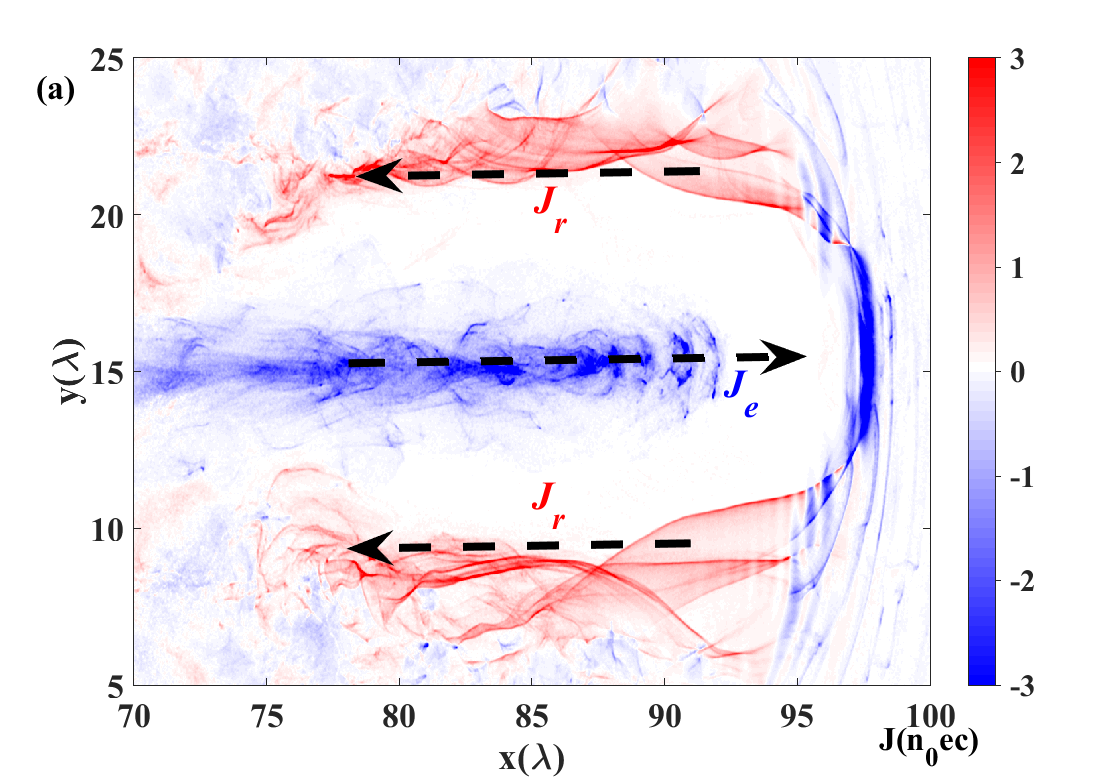}}
\resizebox{90mm}{!}{\includegraphics{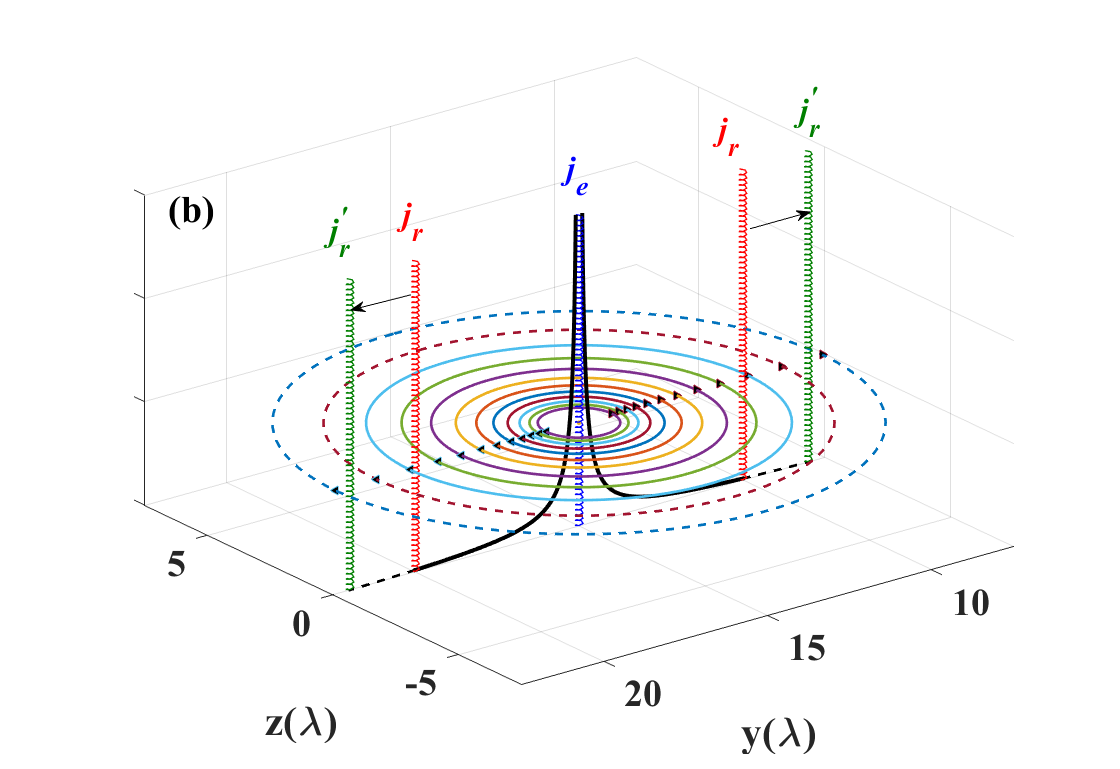}}
\resizebox{56mm}{!}{\includegraphics{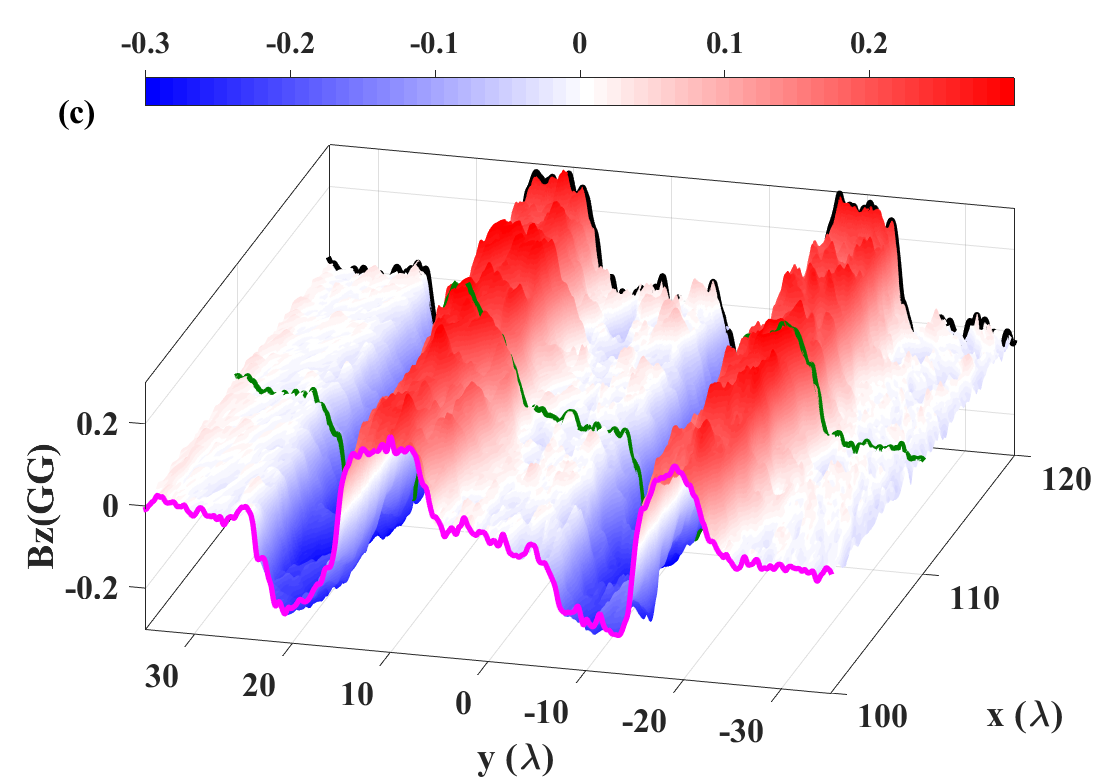}}
\resizebox{56mm}{!}{\includegraphics{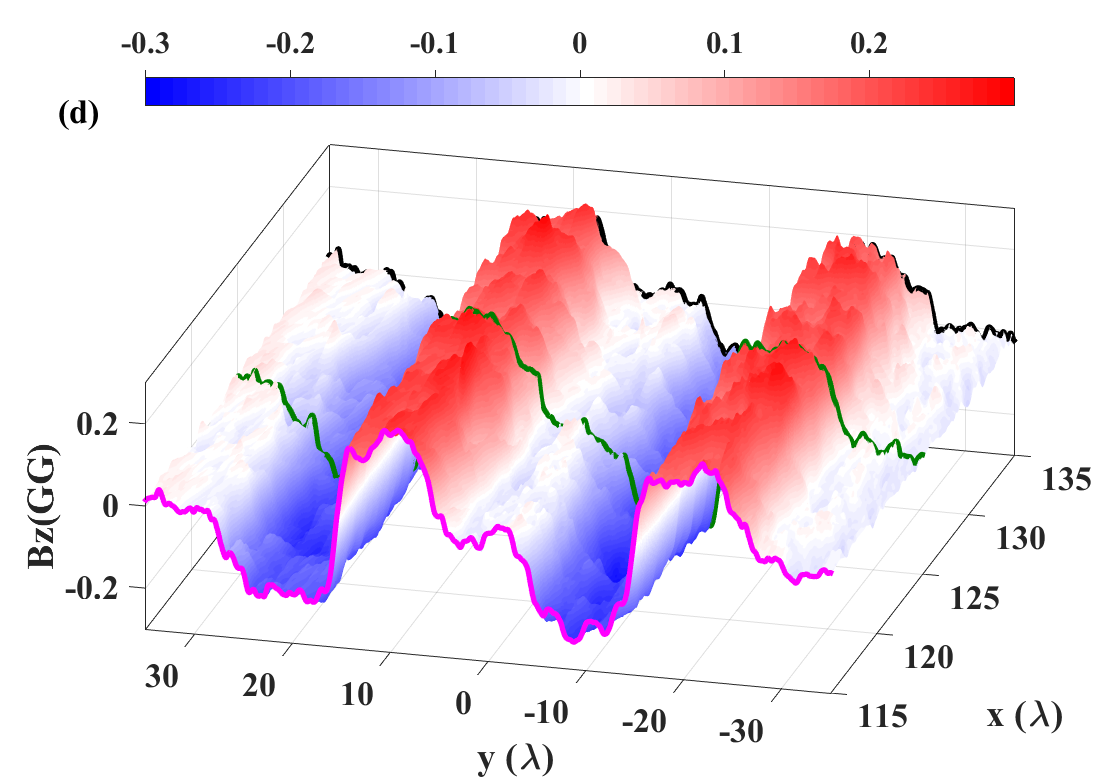}}
\resizebox{56mm}{!}{\includegraphics{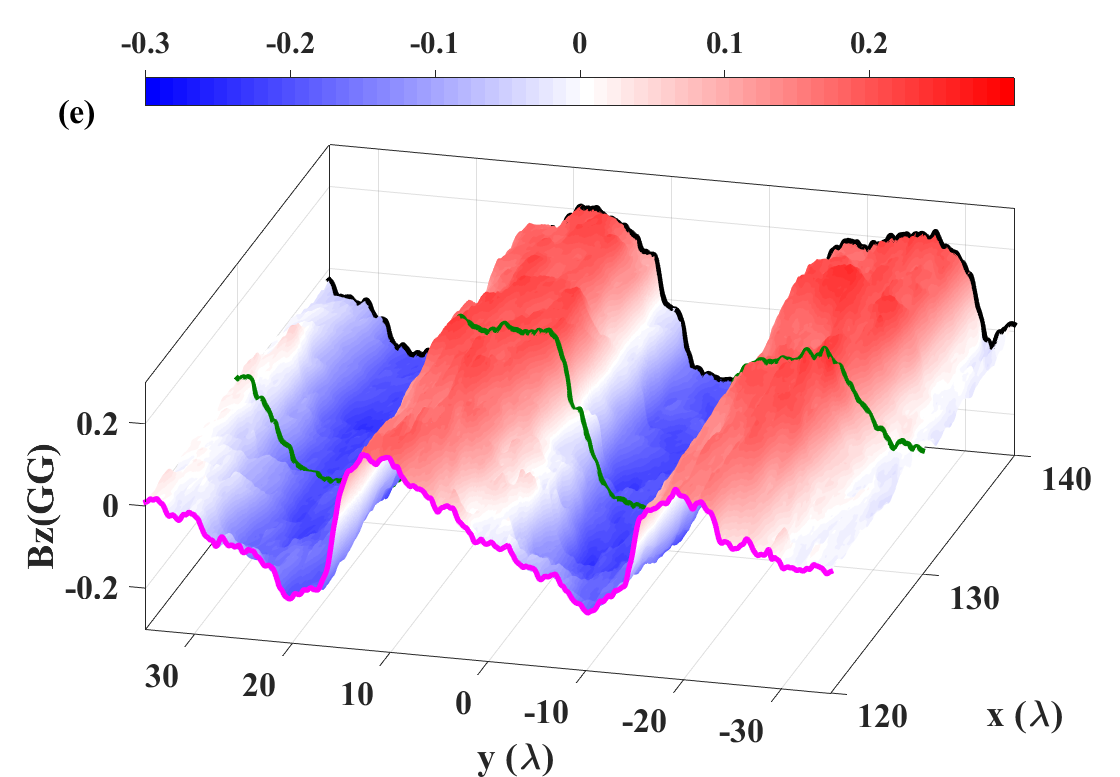}}
\caption{
(color online)
(a) The current density distribution (z=0 plane) at $t=120T_0$ is shown. The dashed arrows indicate the direction of the electron motion. The solid arrows show the radius of the channel structure. $j_e$ and $j_r$ represent the forward and return currents, respectively. (b) The schematic of the magnetic field expansion. (c) to (e) are the surfaces of the magnetic field in the $z=0$ plane.
}
\label{Fig2}
\end{center}
\end{figure*}

When the two regions with opposite polarity magnetic field merge, the so-called X-line structure in MR is formed. Fig. 3(a) depicts the longitudinal currents and the corresponding magnetic field lines in the $(y,z)$ plane at $x=152~\lambda$ at $t=198~T_0$. The vectors indicate the direction of the magnetic field lines. Around the two centers of the accelerated electron beams, the magnetic field loops independently belong to the left and right currents respectively. The separatrix surface forms in the transverse cross-section an eight-like curve with a zeroth X-line locating in the middle of the two currents. Zooming into the region of the X-line in Fig. 3(b), the magnetic field lines have the hyperbolic structure in its vicinity as indicated by the green dashed line. This field can be expressed in terms of the local vector potential as: $A\approx(y^2-b^2x^2)$. Here $b$ is the coefficient which describes the different expansion distance in $y$ and $z$ direction. The magnetic field expand freely in the $z$ direction. However the expansion in $y$ direction is limited by the distance between the two currents. The X-line is the place where redistribution of magnetic fluxes occurs, which changes the connectivity of field lines. Along the separator, a strong electron current is formed. This thin and wide current sheet is one of the signatures of MR due to the particle acceleration via the variation of the magnetic field, which will be discussed in Section V. From the transverse current distribution shown in Fig. 3(c) and (d), one can find the two channels are still expanding to merge. This expansion pushes more magnetic field lines to annihilate and reconnect in the current sheet region.

\begin{figure*}
\begin{center}
\resizebox{80mm}{!}{\includegraphics{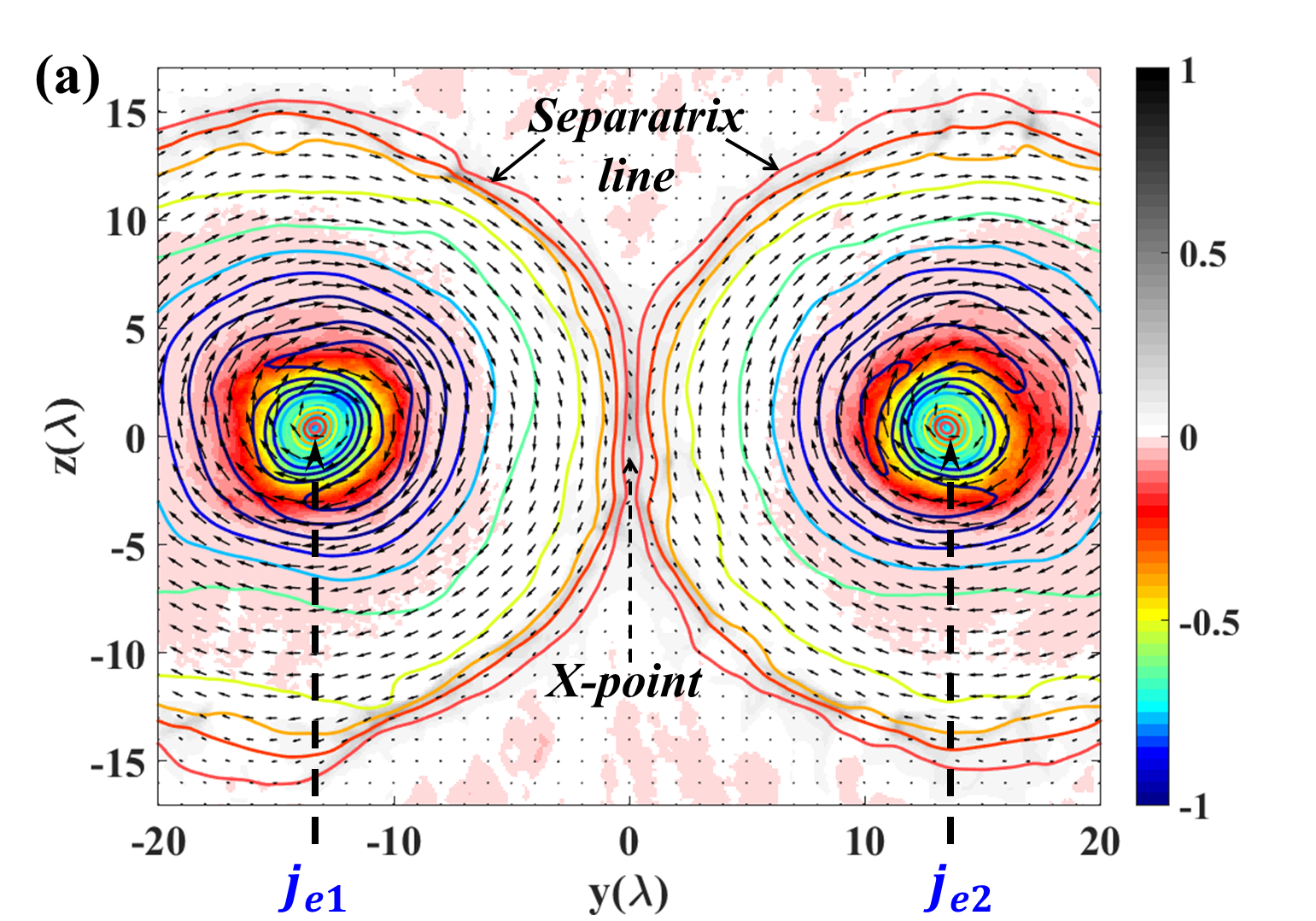}}
\resizebox{80mm}{!}{\includegraphics{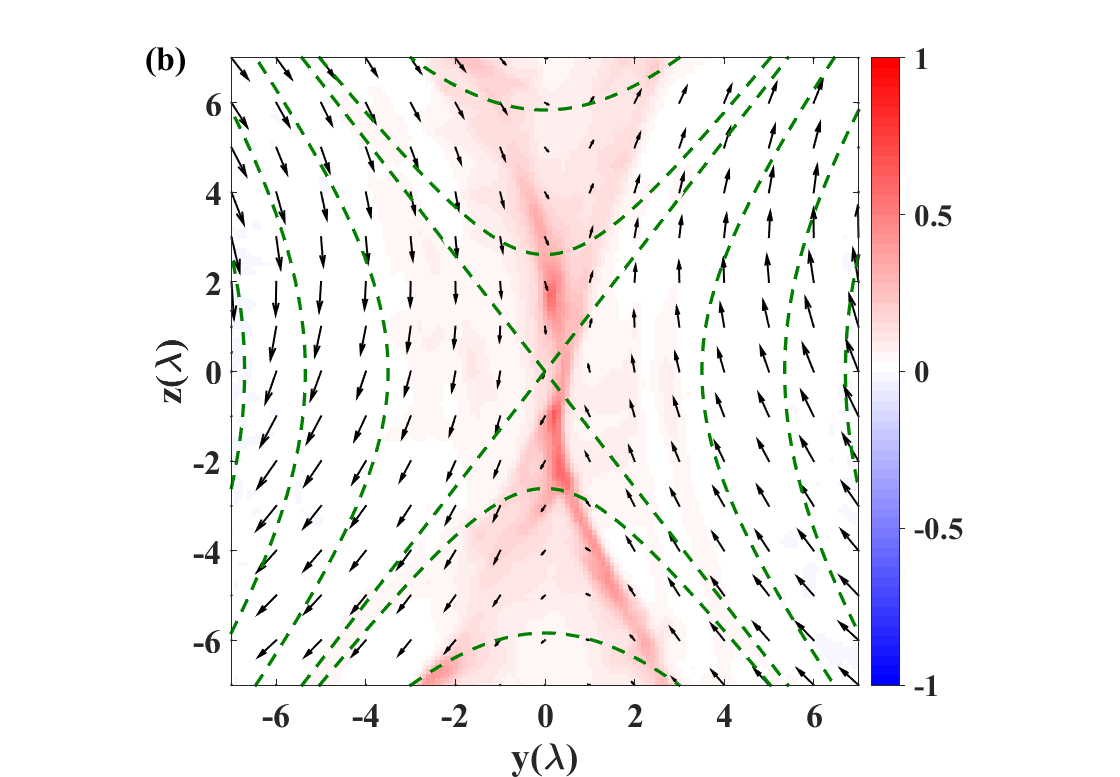}}
\resizebox{80mm}{!}{\includegraphics{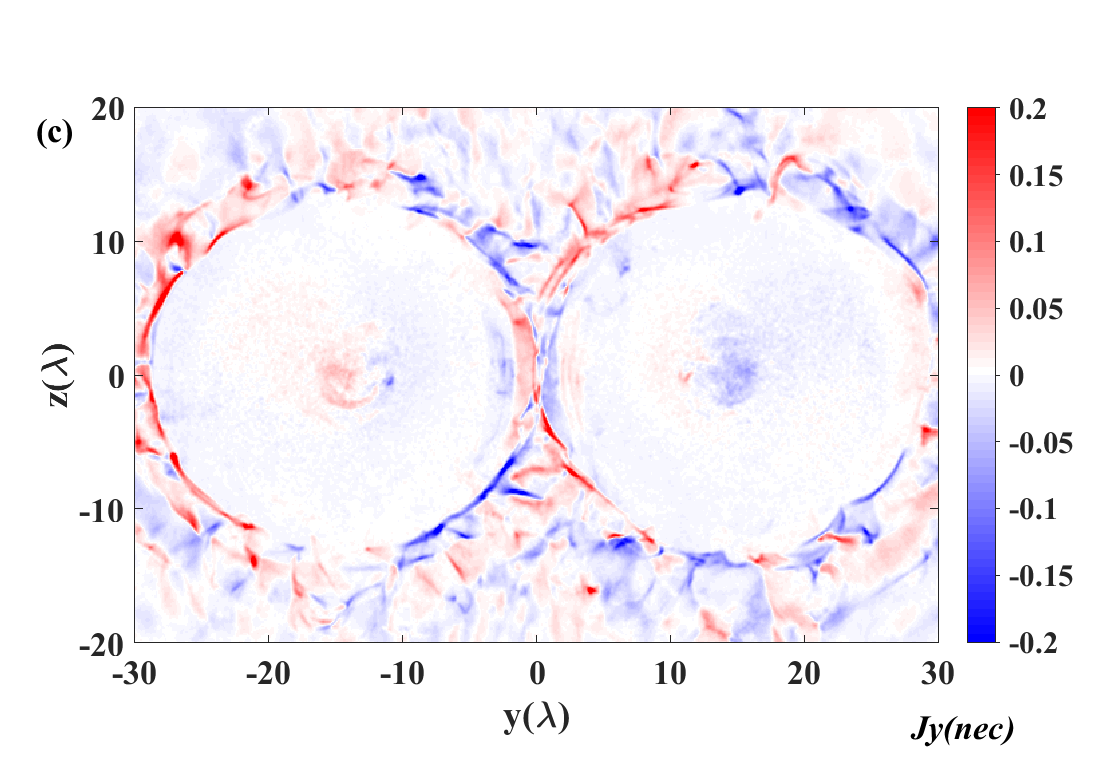}}
\resizebox{80mm}{!}{\includegraphics{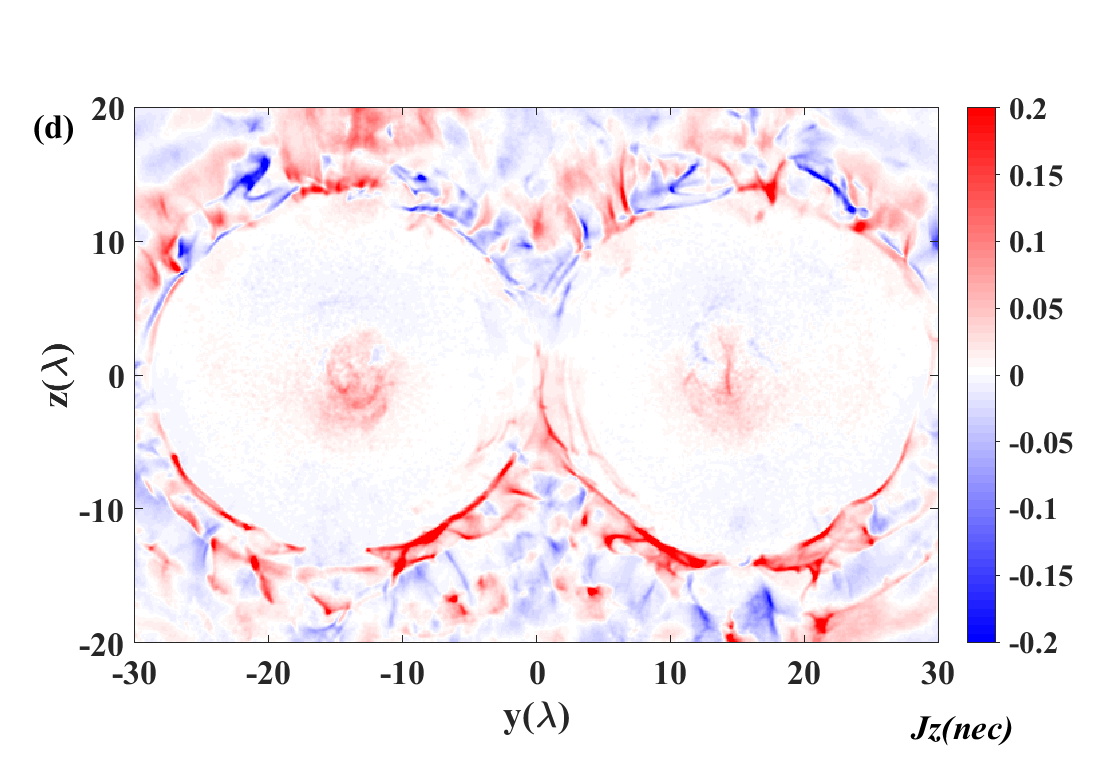}}
%\resizebox{56mm}{!}{\includegraphics{3DBzPro0098.png}}
\caption{
(color online)
(a) The curves are the contour magnetic field strength and the vectors represent the magnetic field lines with directions. The background is the current density distributions in the plane of $x=155\lambda$ at $t=198~T_0$. (b) Zooming to the zeroth-point region in (a). The dashed green lines represents the hyperbolic structure. (c) and (d) are the distributions of $j_y$ and $j_z$.
}
\label{Fig2}
\end{center}
\end{figure*}

\section{V. Electromagnetic Burst and Particle Acceleration}

As mentioned in the previous section, the current sheet is formed due to the particle acceleration in the region where the magnetic fields with opposite polarities annihilate. The particle acceleration in the X-line of MR has been interpreted by several mechanisms including fast shock wave \cite{TsunetaShock}, second type Fermi acceleration by turbulence \cite{MillerFermi} and the reflection in magnetic islands \cite{DrakeMagneticIsland}. Here we proposed the regime of dynamic dissipation of the magnetic field and electromagnetic burst in the ultra-relativistic regime. Recalling the Ampere-Maxwell law~(\ref{eq:Mxwll}), we see that
%$c\nabla \times \mathbf{B}=4\pi\mathbf{j+ j_D}$.
the variation of the magnetic field is sustained by the conduction current with the current density equal to ${\bf j}_e$ and the displacement current  with the density ${\bf j}_D=\partial_t {\bf E}$. Due to the relativistic constraint on the particle velocity, the conduction current density has its limit as $|{\bf j}_e\leq |enc$, where $n$ is the electron density of the current sheet. The local plasma density is relatively low in the downramp region of the target so that the variation of the magnetic field cannot be balanced by only the conduction current. In this case, the displacement current come to play an important role, which is actually the growth rate of the electric field (as noted above, ${\bf j}_D=\partial_t \mathbf{E}$). The magnetic field distribution on the $z=0$ surface is presented in Fig. 4(a) for time at $204~T_0$. The contributions of different terms in Ampere-Maxwell law along the current sheet ($y=0$) are plotted. In the region of $158\lambda<x<170\lambda$, the conduction current ($<{\bf j}_e>_x$, green) is almost negligible and the variation of the magnetic field ($<c\nabla \times \mathbf{B}/4\pi>_x$, black) is identical with the displacement current ($<{\bf j}_D=\partial_t \mathbf{E}>_x$, red). With the increase of the displacement current, a strong longitudinal electric field (blue) is induced. The longitudinal electric field grows as a result of the onset of the displacement current. The peak of the displacement current propagates along the x-direction. Therefore the longitudinal electric field has its peak value behind the summit of the displacement current.
This inductive electric field accelerates the electrons in the backward direction. The magnetic field energy is then transferred to the kinetic energy therefore it is called as dynamic dissipation by S. I. Syrovatskii.

The dynamics of the particles in the vicinity of the X-line has been discussed in Ref.~\cite{BulanovSasorov1976}. Here we employ the main conclusions suitable for our case. The electric and magnetic field around the reconnection region can be approximately described as:
\begin{equation}
\mathbf{E}=E_0\mathbf{\hat{e}_x},
\label{Eq1}
\end{equation}
\begin{equation}
\mathbf{H}=h(-y\mathbf{\hat{e}_z}-z\mathbf{\hat{e}_y}).
\label{Eq1a}
\end{equation}
Then the dynamic equations of the particle acceleration are:
\begin{equation}
\dot{p_x}=-eE_0+\frac{eh}{c}(\dot{y}y-\dot{z}z),
\end{equation}
\begin{equation}
\dot{p_y}=-\frac{eh}{c}\dot{x}y,
\end{equation}
\begin{equation}
\dot{p_z}=\frac{eh}{c}\dot{x}z.
\label{Eq2}
\end{equation}

In the ultra-relativistic case, the scale-length (Larmor radius) characterizing the electron trajectory is $R=cp/(ehr)=E_0/h$ and the characteristic time of electron passing through the X-line region is $T=E_0/ch$. With R and T being the size and time scale related to the non-adiabatic region in the vicinity of the X-line, where the charged particles are not magnetized. Within this region the electron trajectory is given by the solutions of Eq. (4-6) which read as
%Then the solution for Eq. (2) in this case is given by
%
\begin{equation}
\label{Eq3a}
x(t)=ct,
\end{equation}
\begin{equation}
\label{Eq3b}
y(t)=y_0J_0\left(\sqrt{\frac{4hct}{E_0}}\right),
\end{equation}
\begin{equation}
z(t)=z_0I_0\left(\sqrt{\frac{4hct}{E_0}}\right).
\label{Eq3c}
\end{equation}
Here $J_0(x)$ and $I_0(x)$ are the ordinary and the modified Bessel function of zeroth order. From Eq. (\ref{Eq3a},\ref{Eq3b}, \ref{Eq3c}), one can see the electron trajectories oscillate in the $y$-direction and exponentially expand in the $z$-direction. The typical ejected electrons accelerated by the inductive electric field through the vicinity of X-line are selected and the corresponding real trajectories are plotted in Fig. 4(b). The trajectories have clear one period oscillation in the $y$-direction and quick expansion in the $z$-direction, which is well consistent with the theoretical  description. The bottom surface of Fig. 4(b) shows the longitudinal electric field distribution in the plane of $x=165$ at $t=210T_0$. The curves are the contour of the magnetic field strength. The bright spot in the center shows the strong inductive electric field induced by MR, where the magnetic field is almost zero. The strength of the inductive electric field reaches about $\mathrm{E_0\approx30GV/cm}$. The inductive electric field region moves in the forward direction with the propagation of the opposite magnetic polarities. In this case, the backward accelerated electrons only experience a short range in the acceleration phase and then are ejected away from the field. Therefore the energy of the ejected electrons can be estimated as $\mathrm{\cal{E}}_{ke}\sim eE_0\delta l$, here $\delta l$ is the distance in which the electron experiences the field and can be approximately equal to the size of the X-line region as $10\lambda$. Then the characteristic energy is about $\mathrm{\cal{E}}_{ke}\sim eE_0\delta l\approx30~MeV$. Here the energy spectrum of the electrons which are initially localized in the current sheet is provided in Fig. 4(c). An energy peak appears at around $30~MeV$ which is expected by our estimation. The mono-energetic radiation in the astrophysics is one of the difficulties in explaining by other acceleration mechanisms. Here we found by the MR induced particle acceleration, it is natural to obtain the mono-energetic beam.

\begin{figure*}
\begin{center}
\resizebox{80mm}{!}{\includegraphics{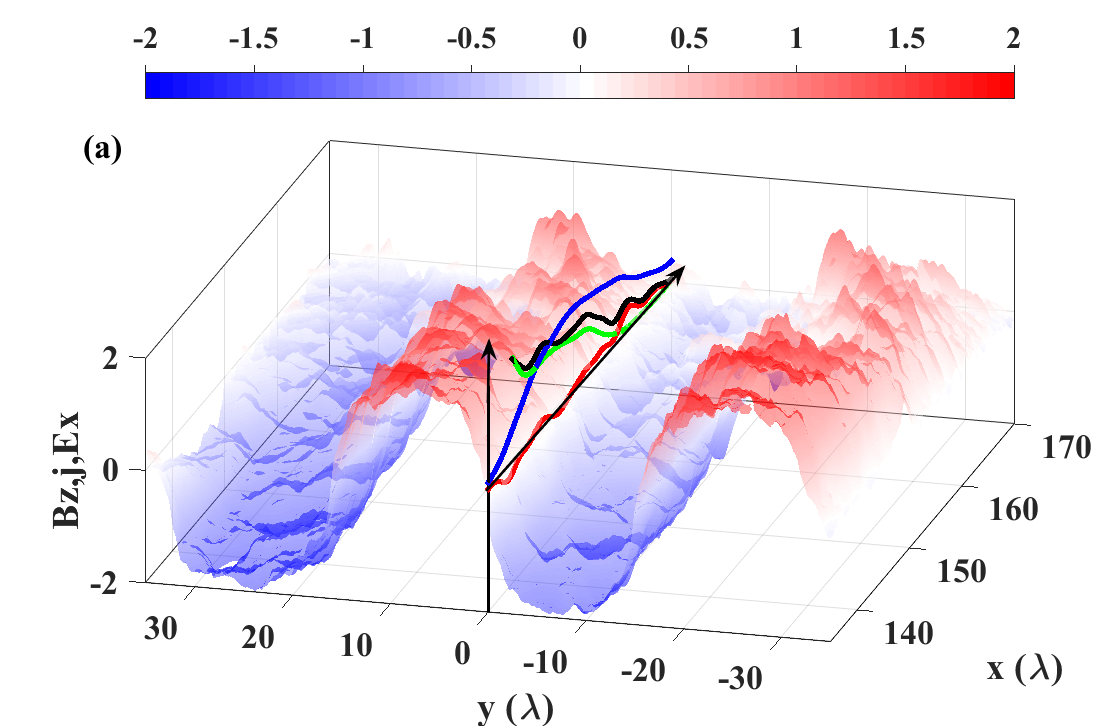}}
\resizebox{80mm}{!}{\includegraphics{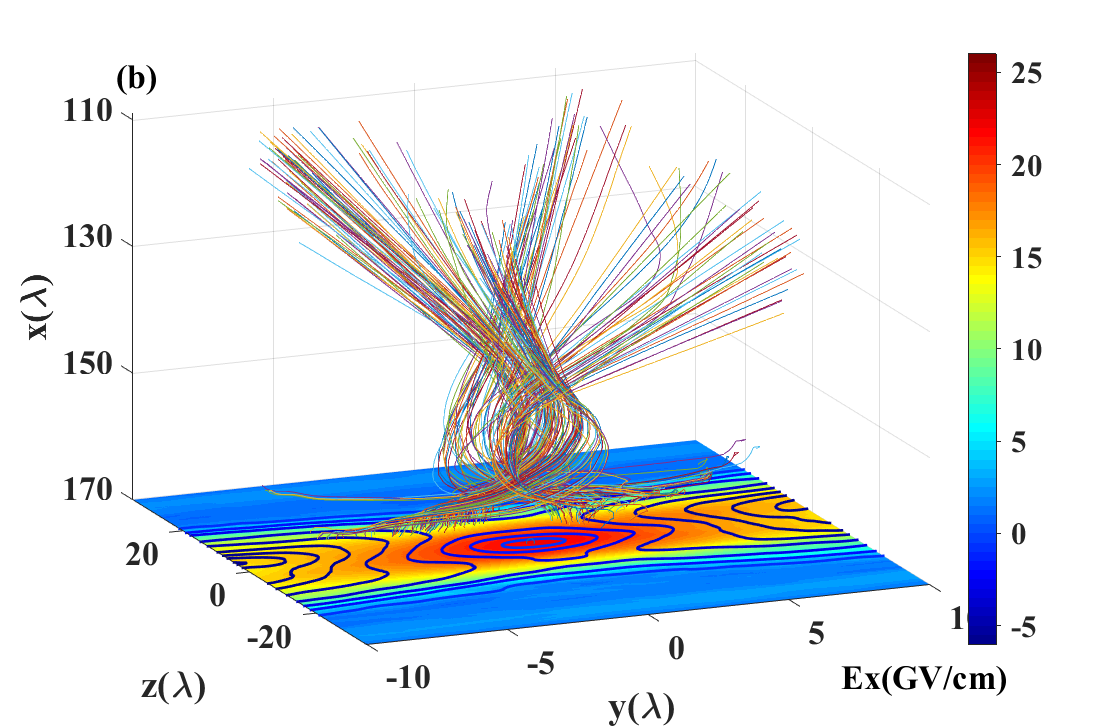}}
\resizebox{80mm}{!}{\includegraphics{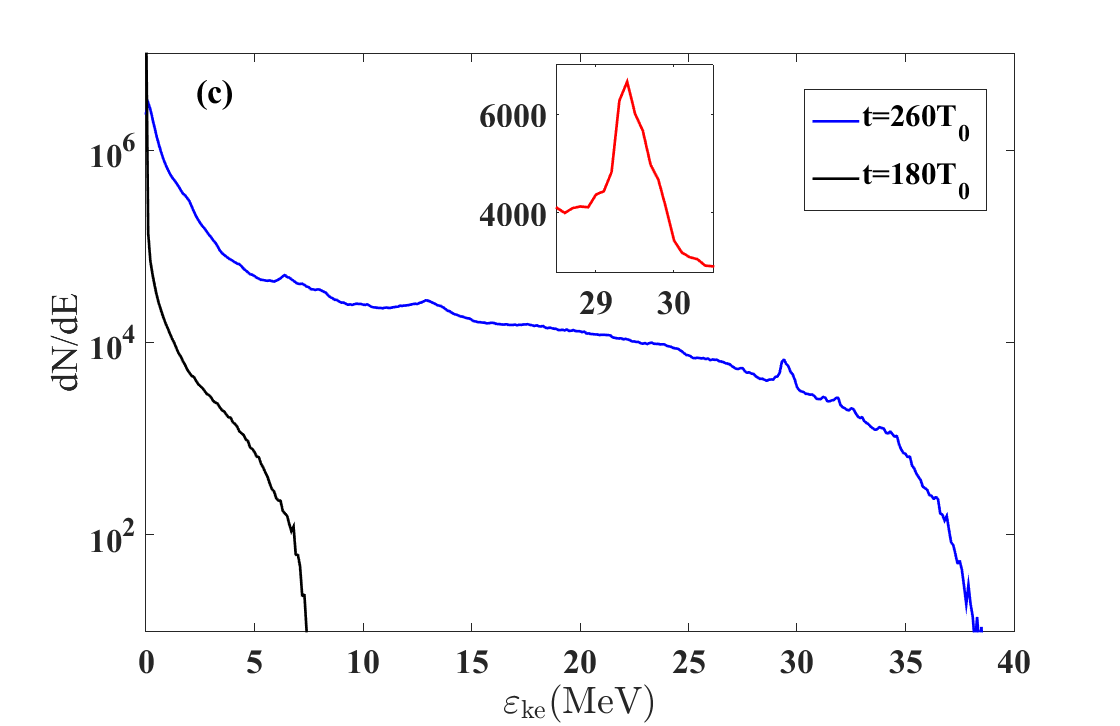}}
\caption{
(color online)
(a) The surface represents the $B_z$ distribution at $t=204T_0$ in the $z=0$ plane. The terms in the Ampere-Maxwell law along $y=0$ are plotted as the curves. $<c\nabla \times \mathbf{B}/4\pi>_x$ (black), $<j_e>_x$ (green) and $<j_D=\partial_t \mathbf{E}>_x$ (red) are normalized to $en_0c$. The longitudinal electric field $<E_x>$ (blue) is normalized to $m_e\omega_0c/e$. All the values are transversely averaged inside the current sheet ($-\lambda<y<\lambda$). (b) The real trajectories of the electrons ejected from the vicinity of X-line (from $185T_0$ to $260T_0$). The distribution of $E_x$ in the plane of $x=165$ at $t=210T_0$ is on the bottom. The curves in the bottom plane represent the contour of magnetic field strength. (c) The energy spectra of the electrons initially locating in the current sheet region. The inset shows the quasi-mono-energetic electrons spectrum.
}
\label{Fig3}
\end{center}
\end{figure*}

\section{VI. Pattern of the Longitudinal Magnetic Field Induced by the Hall effect}

 The Hall effect manifests  the basic properties of the magnetic reconnection  in collisionless plasmas (e. g. see Refs. \cite{BPS, KA98, 0741-3335-59-1-014029}).
 The Hall effect is closely related to the in the transverse electric field and current excitation, seen in the quadruple magnetic field formed in the vicinity of the X-line. With the magnetic field annihilation and the inductive field increasing, it generates in-plane currents due to the decoupling between electrons and ions. The plasma transversely drifts according to the distribution of the magnetic fields and the inductive electric field with the drift velocity $\mathrm{\textbf{v}_d=c(\textbf{E}\times\textbf{B})/B^2}$ as shown in Fig. 5(a). In the vicinity of the magnetic null line, the currents generate the longitudinal magnetic field $B_x$ with the feature of a characteristic quadrupole. It is also such the $B_z$ quadruple patterns are distinctly seen in Fig. 5(b).

\begin{figure*}
\begin{center}
\resizebox{80mm}{!}{\includegraphics{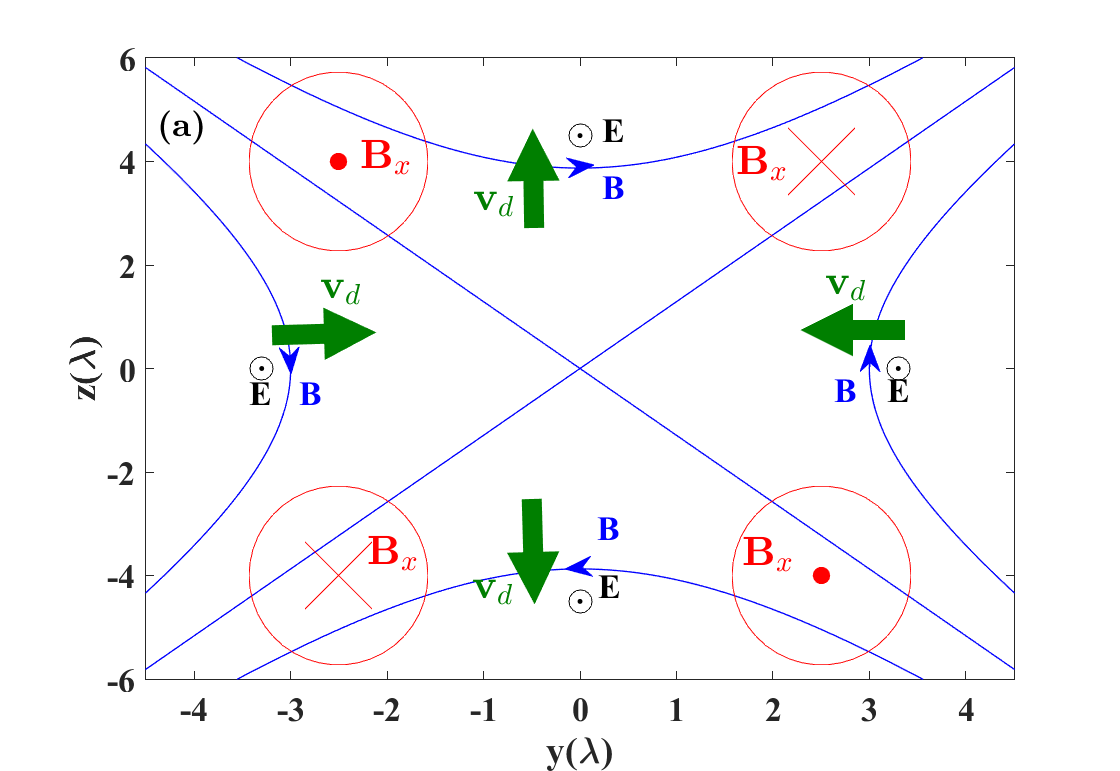}}
\resizebox{80mm}{!}{\includegraphics{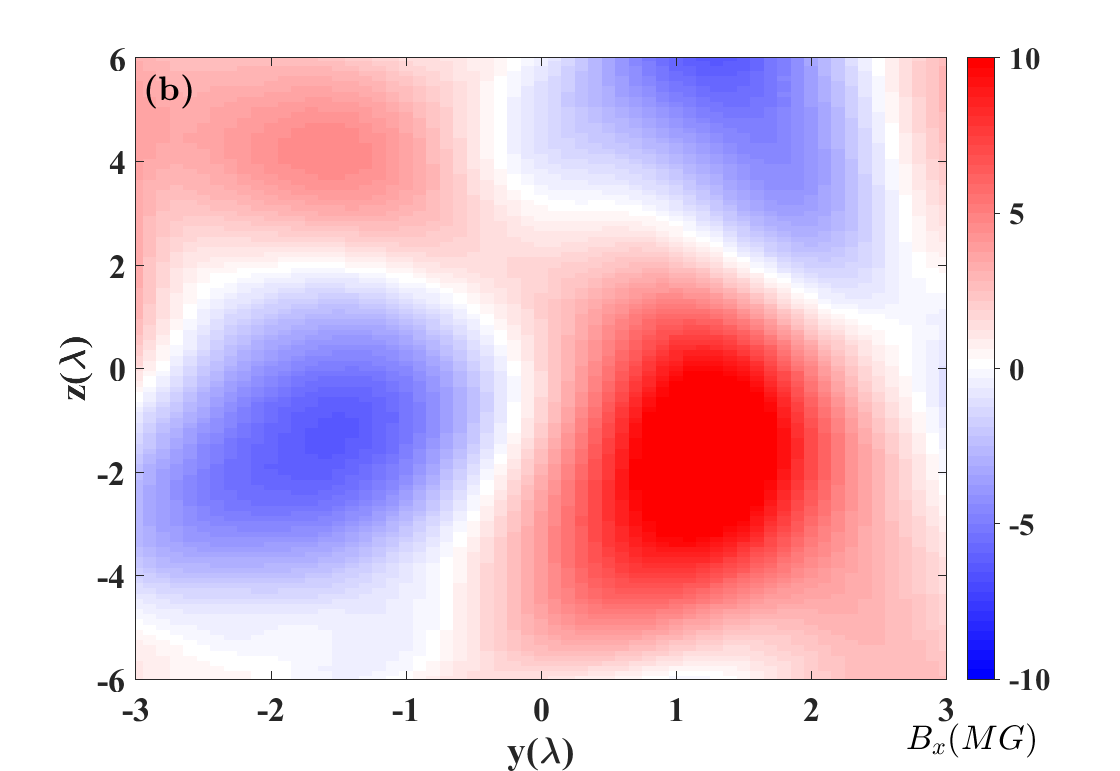}}
\caption{
(color online)
(a) The schematic of the field and drift distribution in the vicinity of the magnetic null point. (b) The longitudinal magnetic field ($B_x$) distribution in the plane of $x\sim 160\lambda$ at $t=198T_0$.
}
\label{Fig3}
\end{center}
\end{figure*}

\section{VII. Tearing-like Mode Instability in Current Sheet and Magnetic Islands}

The current sheet is unstable and it may break-up into
filaments due to the development of the tearing mode instability.
The corresponding schematics are shown in Fig. 6(a) to (c).
In Figs. 6(d) to (f), we present the evolution of the current density in the $(y,z)$
plane with the magnetic field lines in the 3D simulation results. The snapshots correspond to the time at $200~T_0$,
$202~T_0$ and $206~T_0$. The current sheets clearly extend in the $z$-direction accompanying with filamentation.
The electrons accelerated by the inductive electric
field via magnetic field annihilation generate a current in the X-line region as shown in Fig. 6(a) and (d).
The trajectories of the accelerated electrons experience an expansion in the $z$-direction as discussed in Section V. The
corresponding magnetic field produced by this current changes the local field topology. The X-line now splits into two
symmetric zeroth lines in the $z$-direction as $X^{\prime}$ in Fig. 6(b). This transverse expansion continuous and
finally form a thin but wide current sheet. Accompany with the current sheet formation, the tearing mode instability leads
to the current filamentation and pinching.
As shown in Fig. 6 (c) and (f), the current breaks into separated pieces and several magnetic islands are formed in the current sheet.
The filamentation and the breaking also consistent with the electron trajectories shown in Fig. 4(b).

\begin{figure*}
\begin{center}
%\resizebox{56mm}{!}{\includegraphics{Tear1.png}}
%\resizebox{56mm}{!}{\includegraphics{Tear2.png}}
%\resizebox{56mm}{!}{\includegraphics{Tear3.png}}
%\resizebox{56mm}{!}{\includegraphics{CurretnXBVCX150To160T0072.png}}
%\resizebox{56mm}{!}{\includegraphics{CurretnXBVCX150To160T0074.png}}
%\resizebox{56mm}{!}{\includegraphics{CurretnXBVCX150To160T0076.png}}
\resizebox{180mm}{!}{\includegraphics{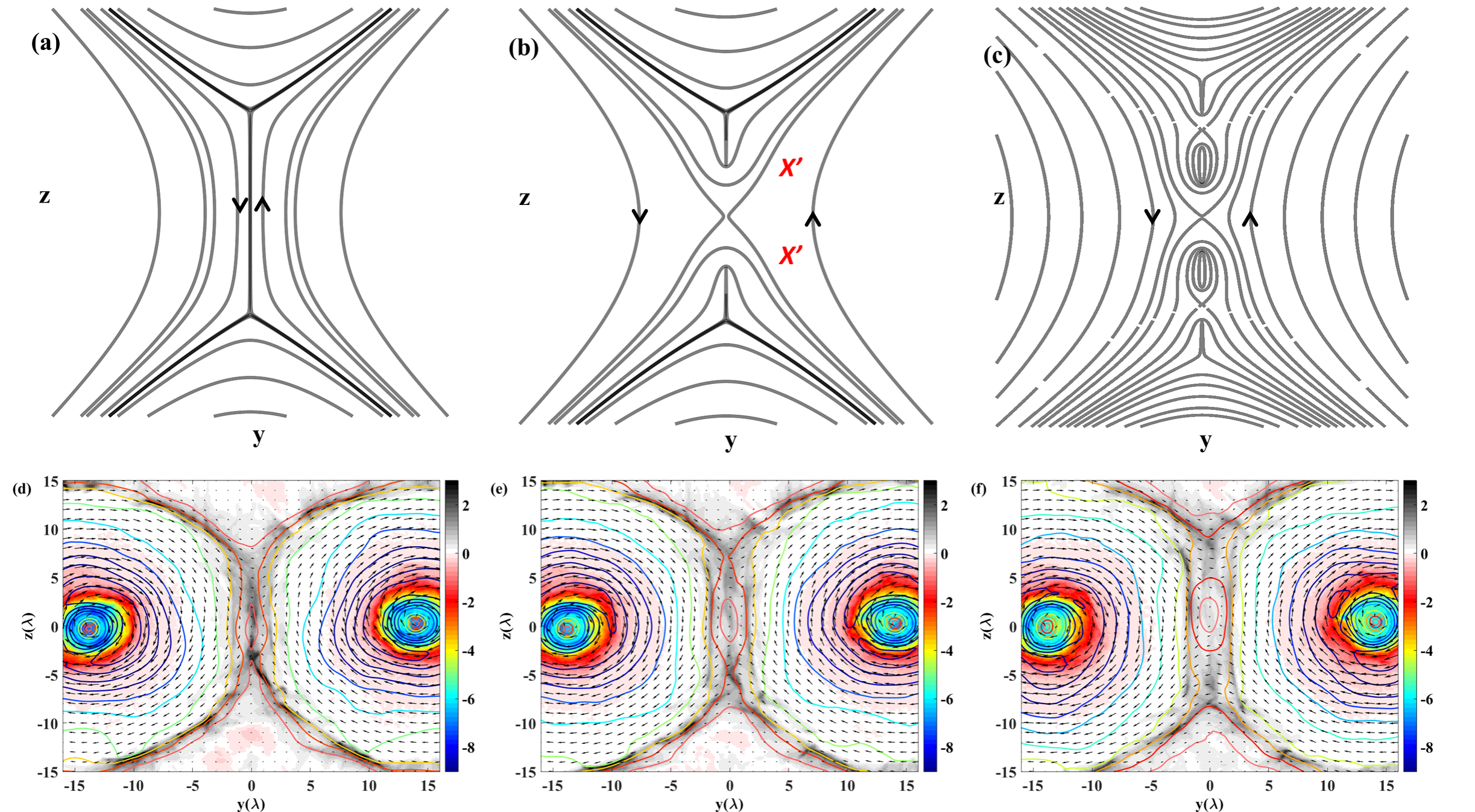}}
\caption{
(color online)
(a) to (c) are the constant vector potential surfaces corresponding to the magnetic field in the vicinity of the current sheet presenting the current sheet expansion and the bifurcation of the X-line. (d) to (f) are the results from the 3D simulations. The corresponding times are $200~T_0$, $202~T_0$ and $206~T_0$. The curves are the contour magnetic field strength and the vectors represent the magnetic field lines with directions. The background is the current density distributions in the plane of $x=155\lambda$.
}
\label{Fig3}
\end{center}
\end{figure*}

%In conclusion, we demonstrate an approach towards controlling the relativistic magnetic reconnection by using a solid cone target to split the laser pulse. Accelerated electron beams are generated in the subsequent gas jet plasma, which forms the strong magnetic fields. In both 2D and 3D PIC simulations, the high energy backward moving electrons form the thin current sheets which are typical for magnetic reconnection. It is also proved the conversion of the magnetic field energy into particle kinetic energy via the displacement current and the growth of the induced electric field. The energy of accelerated electrons reaches hundreds MeV. Although it is lower than the electrons energy acceleration in other regime like by the wakefield, it is an importnat signature of magnetic reconnection. Using the solid cone target enables perfect synchronizing of the parallel splitting pulses. This would be difficult to realize by several pulses. The setup is easy for experiments for the upcoming laser facilities \cite{ELIWhitebook}.

\section{VIII. Discussions and Conclusions}

In conclusion, we investigate the magnetic reconnection driven by laser-plasma interaction by using the 3D kinetic simulations. It presents the formation and evolution of the current sheet in a collisionless plasma during magnetic field annihilation in ultra-relativistic limit.
The accelerated electron beams generated in the gas jet plasma create strong magnetic fields. The annihilation of opposite magnetic polarities is accompanied by an electromagnetic burst generation whose strong non-stationary electric field accelerates the charged particles within the current sheet. It is found that the displacement current plays important role in the ultra-relativistic MR to induce the significant growth of the longitudinal electric field. In the vicinity of the magnetic null line, charged particle acceleration is observed. Since the inductive electric field moves in the forward direction withrespect to  the propagating of the laser field, the electrons, which are accelerated in the backward direction, experience only an instantaneous kick. Therefore the corresponding electron bunch has a relative small energy spread. The dynamics of the particles accelerated by this field in the region in the vicinity of the X-line has been studied. Narrow energy spectrum electron beam is obtained which will be useful in explaining the radiation spectrum obtained in the astrophysics. One of the intriguing problems standing in astrophysics for a number of years is explanation of the detected gamma-ray spectrum pointing towards a very narrow particle spectrum, which is one of the arguments against the shock-acceleration regime. Our results of the mono-energetic bunch generation provide a clear signature to support the particle acceleration via MR regime. Due to the development of the tearing-like mode instability, the current sheet breaks into separated pieces. It leads to formation and evolution of the magnetic islands in the current sheet. The requirement of the laser energy can be expected to be fulfilled by the upcoming facilities like ELI-Beamlines \cite{ELIWhitebook}. The regime proposed can be used for formulating the program of forthcoming experiments, including the research in laboratory astrophysics \cite{Bulanov2009, 0741-3335-59-1-014029, BulanovPPR2015}.

~\

\begin{acknowledgments}
We thank Prof. M. Tavani for fruitful discussions.
This work was supported by the project ELITAS (CZ.02.1.01/0.0/0.0/16\_013/0001793) and by the project High Field Initiative (CZ.02.1.01/0.0/0.0/15\_003/0000449) from European Regional Development Fund. Computational resources were provided by the MetaCentrum under the program LM2010005, IT4Innovations Centre of Excellence under projects CZ.1.05/1.1.00/02.0070 and LM2011033 and by ECLIPSE cluster of ELI-Beamlines. This work was also supported by PRESTO (JPMJPR15PD) commissioned by Japan Science and Technology Agency. The EPOCH code was developed as part of the UK EPSRC funded projects EP/G054940/1.
\end{acknowledgments}

%\bibliography{YJGu_MRSPILT}

\end{document}